\documentclass[11pt]{article}
\usepackage[english]{babel}
\usepackage[utf8x]{inputenc}
\usepackage{amsmath}
\usepackage{graphicx}
\usepackage{etoolbox}
\usepackage{changepage}
\usepackage{titlesec}
\usepackage[parfill]{parskip}
\usepackage[margin=1in]{geometry}
\usepackage{times}
\usepackage{float}
\usepackage{lipsum} 
\usepackage{setspace}
\usepackage{amsmath}
\usepackage[flushleft]{threeparttable}
\usepackage{makecell,booktabs}
\usepackage[numbers,super]{natbib}
\usepackage{censor}
\usepackage{caption}
\usepackage{subcaption}
\usepackage{amssymb,verbatim,epsfig}
\usepackage{amsmath,setspace}
\usepackage[usenames,dvipsnames]{color}
\usepackage{url}
\usepackage{amssymb, amsmath, verbatim, epsfig}
\usepackage{setspace}
\usepackage{enumitem}

\newcommand{\fig}[1]{{\bf Figure \ref{#1}}}

\usepackage{amssymb, amsmath, verbatim, epsfig}
\usepackage{xcolor}

\titleformat*{\section}{\normalsize\bfseries} 

\title{\vspace{-1in}\large \textbf{Utilizing Online $\&$ Open-Source Machine Learning Toolkits to Leverage the Future of Sustainable Engineering}} 
\date{} 

\title{\bf 
\vspace{-1in}
Utilizing Online $\&$ Open-Source Machine Learning Toolkits to Leverage the Future of Sustainable Engineering}

\author{Andrew Schulz$^{1,+,*}$, Suzanne Stathatos$^{2,+}$, Cassandra Shriver$^{3}$, Roxanne Moore$^{1}$ \\
\text{\small{Schools of Mechanical Engineering$^1$ and Biological Sciences$^3$}}\\
\text{\small{Georgia Institute of Technology, Atlanta, GA 30332, USA}} \\
\text{\small{School of Computing and Mathematical Sciences$^2$}} \\
\text{\small{California Institute of Technology, Pasadena, CA 91125, USA}}\\
\text{\small{}}}

\begin{document}
 \maketitle

 \noindent + designates co-first author \\
 \noindent * designates the corresponding author

\section*{Abstract}
The United Nations Sustainable Development Goals (SDGs) have become a foundational metric for advancing engineering education in non-traditional ways, similar to the NSF’s Big 10 Ideas and the Grand Challenges. Recently, there has also been a national push to use machine learning (ML) and artificial intelligence (AI) to advance engineering techniques in all disciplines ranging from advanced fracture mechanics in materials science to soil and water quality testing in the civil and environmental engineering fields. Using AI, specifically machine learning, engineers can automate and decrease the processing or human labeling time while maintaining statistical repeatability via trained models and sensors. Edge Impulse has designed an open-source TinyML-enabled Arduino education tool kit for engineering disciplines. This paper discusses the various applications and approaches engineering educators have taken to utilize ML toolkits in the classroom. We provide in-depth implementation guides and associated learning outcomes focused on the Environmental Engineering Classroom. We discuss five specific examples of four standard Environmental Engineering courses for freshman and junior-level engineering. There are currently few programs in the nation that utilize machine learning toolkits to prepare the next generation of ML $\&$ AI-educated engineers for industry and academic careers. This paper will guide educators to design and implement ML/AI into engineering curricula (without a specific AI or ML focus within the course) using simple, cheap, and open-source tools and technological aid from an online platform in collaboration with Edge Impulse. Specific examples include 1) facial recognition technologies and the biases involved, 2) air quality detection using an accelerometer, 3) roadside litter detector, 4) automated bird identifier, and 5) wildlife camera trap detection.

\section{Introduction}
In 2015, while seeking to create a global development framework, the United Nations (UN) formulated the seventeen interrelated UN Sustainability Development Goals (SDGs). The SDGs include 
fighting poverty and hunger, and promoting health, quality education, gender equality, clean water and sanitation, affordable and clean energy, economic growth, infrastructure, and innovation in the industry, reduced inequalities, sustainable communities, responsible consumption, climate action, marine and terrestrial life, peaceful, strong, and just institutions, and global partnerships to meet the goals.~\cite{unsdgs}. The SDGs emphasize interconnected socioeconomic, environmental, and political aspects of sustainable development and encourage collaboration and partnership between groups working toward their defined goals. The SDGs specifically emphasize how civil and environmental engineering are crucial to meet their goals, and it has become increasingly apparent ~\cite{sdgs2020} that artificial intelligence (AI) has become a key component to these reach goals. We first outline how we suggest civil and environmental engineering curricula should be (re)formulated to meet the SDGs and then spend the majority of the paper discussing AI's impact both within and outside of these engineering disciplines to meet the SDGs.

\subsection{Engineering Education to work toward the SDGs}

\subsubsection{Civil Engineering Curriculum}
Both Civil and Environmental Engineering have direct connections to the SDGs in their future curriculum outlines. To meet the 2030 Agenda outlined by the SDGs for emerging civil engineers, the college curriculum will need to prepare graduates to apply knowledge of mathematics through differential equations, calculus-based physics, chemistry, and at least one additional area of basic science. Students will need to know how to:
\begin{enumerate}
    \item apply probability and statistics to address uncertainty; 
    \item analyze and solve problems in at least four technical areas appropriate to civil engineering;
    \item  conduct experiments in at least two technical areas of civil engineering and analyze and interpret the resulting data; 
    \item design a system, component, or process in at least two civil engineering contexts;
    \item include principles of sustainability in design;
    \item explain basic concepts in project management, business, public policy, and leadership; 
    \item analyze issues in professional ethics;
    \item and explain the importance of professional licensure.
\end{enumerate}
In this paper, we discuss several connections, not just with these SDG connections and Civil Engineering. Still, we especially believe that the case studies of edge computing and machine learning give direct connections to applying probability, analyzing and solving problems, conducting experiments, and designing a process in civil engineering contexts, all discussed above. 

\subsubsection{Environmental Engineering Curriculum}
For emerging environmental engineers to meet the SDGs, students will need to:
\begin{enumerate}
    \item Have hands-on laboratory experiments;
    \item Analyze and interpret data from their experiments in more than one central environmental engineering focus area, e.g., air, water, land, and environmental health.
    \item Design at least one environmental engineering system that includes considerations of risk, uncertainty, sustainability, life-cycle principles, and environmental impacts
\end{enumerate}
These direct connections of environmental engineering influence of the SDGs directly coincide with the civil engineering outcomes listed in the previous section. In the case studies, we discuss several examples of hands-on laboratory experiments that allow students to use novel techniques to understand and interpret data from environmental engineering systems.  

\subsubsection{Artificial Intelligence's influence in engineering}

The internet revolutionized the amount of data accessible to all (Big Data). Applications and websites like Reddit, Twitter, Wikipedia, iNaturalist, and Merlin Bird ID show the rise and regularity of crowdsourced data. The rise of public-accessible machine-learning-based products  like ChatGPT, Dall-e, and iPhone-unlocking facial recognition illustrate the emergence of artificial intelligence (AI) across vast public sectors. AI is predicted to affect global productivity\cite{acemoglu2018}, to promote/expose problems in diversity, equity, and inclusion~\cite{bolukbasi2016}, to impact conservation and biodiversity monitoring~\cite{norouzzadeh2018}, and to increase the ability to do climate monitoring and forecasting ~\cite{sdgs2020}. The 2020 Nature Communication study \cite{sdgs2020} suggests that AI will influence the ability to meet all 17 Sustainability Development Goals (SDGs) set out in their 2030 Agenda ~\cite{un2015}. Vinuesa, et. al.~\cite{sdgs2020} illustrate and discuss how AI can enable or inhibit the 2030 Agenda for the United Nation's SDGs. 

Academia and industry alike will likely incorporate aspects of AI into existing engineering processes. It is critical to know the benefits and detriments of AI as those processes get modified. Academia will, thus, need to incorporate aspects of machine learning and/or data science into its curriculum pipelines to train the next generation of engineers.

\subsection*{What is Machine Learning (and Artificial Intelligence)?}
Let's take a step back and precisely define these fields. Artificial Intelligence is the broad field in which machines are developed to mimic (and exceed) human capabilities. AI encapsulates every aspect of human intelligence so that machines can simulate humans without human interference. \cite{cs221GeneralAISlides} 

Machine learning is a subcategory of artificial intelligence that provides the tools necessary for machines to exhibit human capabilities. These tools include deep learning and neural networks, computer vision, and natural language processing. Machine learning is used in the real world to recognize and categorize emails as spam, translate speech to text, and classify images (i.e., dog images vs. cat images).

By studying and experimenting with machine learning, programmers test the limits of how much a computer system can improve perception, cognition, and performance on a given set of tasks. They do this through algorithm development and data collection. \cite{MostafaLFD2012}

\subsection*{Why is Machine Learning Important in Education}

Machine learning has applications beyond the realm of engineering, as several disciplines, including computer science\cite{de_freitas_im_2021,steinbach_teaching_2021,shouman_experiences_2022,cole_teaching_2023}, physics\cite{lauer_multi-animal_2022}, medicine\cite{mooney_big_2018}, biology and ecology\cite{tuia_perspectives_2022}, and public policy\cite{ackermann_deploying_2018} have all grown in computational techniques by utilizing machine learning for various tools. As applications in these fields expand, so does the need for developing the next generation of machine learning-conscious students at the undergraduate level. There are now several online teaching modules working on machine learning in education. Free, general, machine learning and statistics classes are provided online by Coursera, Udacity, Khan Academy, and more\cite{khanAcademy, coursera, udacity}. Cloud Providers like Amazon AWS, Google GCP, and Microsoft Azure provide free online tutorials with step-by-step guides on how to plug into and use their machine learning services\cite{aws, gcp, azure}. 
These online services are growing partly because they are trying to reduce the gap between those who "know" machine learning technologies and those who use machine learning libraries.

Machine learning's reach has expanded to many application areas, as enumerated above. However, this expanded reach increases demand for machine learning education, particularly for those who do not come from a computer science or statistics-heavy background. While machine learning is still a growing research area, there is limited consensus on teaching it to interdisciplinary or non-computer science audiences. However, Machine Learning Engineering is cited as the fourth largest growing field in the job market.\cite{linkedin_news_linkedin_2022} Machine Learning Engineering is needed for various fields, yet there is a bottleneck in how many people are learning machine learning in higher education settings. Moreover, applying machine learning often requires subject matter experts in those fields to determine if the method or system is generating accurate predictions. For example, machine learning on medical MRI imagery may require MRI experts to label parts of those images so the machine learning model can learn correctly. 

Today, there is a lot of ``over the wall" machine learning \cite{overthewall} where subject matter experts throw their data to machine learning experts who know little to nothing about the subject. They create a model that they think works well and throw it back ``over the wall" to the subject matter experts. This process can be repeated many times and often results in frustration from both sides.

Machine Learning Engineers coming out of engineering programs are in short supply and often less than what is needed by many companies. Instead, people must be familiar with machine learning to set it up and use existing ML tools on their data. 

With the expanded use of machine learning, educators need to work to find new and innovative ways to teach machine learning to undergraduate engineers \cite{weerts_teaching_2022}. One such way would be via a case study. Educators could, for example, teach a class on how to apply machine learning by leveraging conservation tools. In it, they could have students simultaneously implement Edge Impulse devices, learn about the role of sustainability goals in engineering, and learn how to apply machine learning to a specific domain.

Prior work has considered the challenges of teaching machine learning courses to cross-disciplinary audiences such as non-CS undergraduates\cite{sulmont_what_2019}, business students\cite{wunderlich_machine_2021}, artists\cite{fiebrink_machine_2019}, materials scientists\cite{sun_teaching_2022}, biologists \cite{magnano_approachable_2022}, and graduate ecologists\cite{cole_teaching_2023}. One of the chief challenges with non-CS undergraduate students is that they lack a background in advanced programming for machine learning projects. This is especially the case for some fields in the biological sciences. However, many data collection techniques in fields like ecology rely on long field observation records or data analysis in behavioral ecology. These fields would greatly benefit from computer science techniques such as machine learning, algorithms, computer vision, and artificial intelligence to allow large data sets to be analyzed in minutes compared to weeks. 

\subsection*{How does Machine Learning Work, Broadly?}
Broadly, machine learning works by taking in some data, feeding the data through an algorithm, and having the algorithm predict things about the data. The data is typically split into some data to train the machine learning model and separated data to validate that the model works as expected. The data can be labeled or unlabeled. In labeled scenarios, labels could be like a box around something in an image, as shown in the example in \fig{fig:ml_conservation}, a "SPAM" or "NOT SPAM" text categorization, and many more.  The algorithm "learns" by penalizing wrong guesses and encouraging correct guesses iteratively over the data. In the unlabeled scenario, the algorithm learns innate patterns in the data, such as clustering patterns that might be difficult to recognize for high-dimensional data. Currently, there are many novel ways that machine learning is learned in and outside of the classroom. 

\subsection*{How do People Learn Machine Learning Today?}
The primary modes of delivery of machine learning education are: 
\begin{enumerate}
    \item In college settings,
    \item Using MOOCs (i.e., Coursera, Khan Academy, Udacity),
    \item  or Using "Edge" Devices.
\end{enumerate}
Each of these modes of delivery has unique advantages and disadvantages. Items like MOOCs allow the masses to access information but have challenges with techniques like one-on-one assistance and active learning environments. Therefore in this manuscript, we will focus on the ability to use "edge" devices in college setting classrooms as an educational tool. 

\section{Edge Impulse}
\subsection*{What are ``Edge" Devices?}
\textbf{Edge devices} are things we know and love - phones, tablets, laptops, and personal computers. They are meant to distinguish from non-edge devices in a server room or on cloud computers. Much of machine learning today runs on non-edge devices because they have more relaxed storage and computational limits. However, they contribute to the limited accessibility problem of machine learning. Fortunately, edge devices are becoming increasingly common to run machine learning models directly. There are a few different edge devices that can be used as educational tools, including items like EdgeImpulse, Google Coral, or Edge boards from Arduino. We will focus this manuscript on the EdgeImpulse platform for educational purposes, but readers should be aware of other potential devices to use in their classrooms. 

\begin{figure}
\centering
\includegraphics[width=0.7\textwidth]{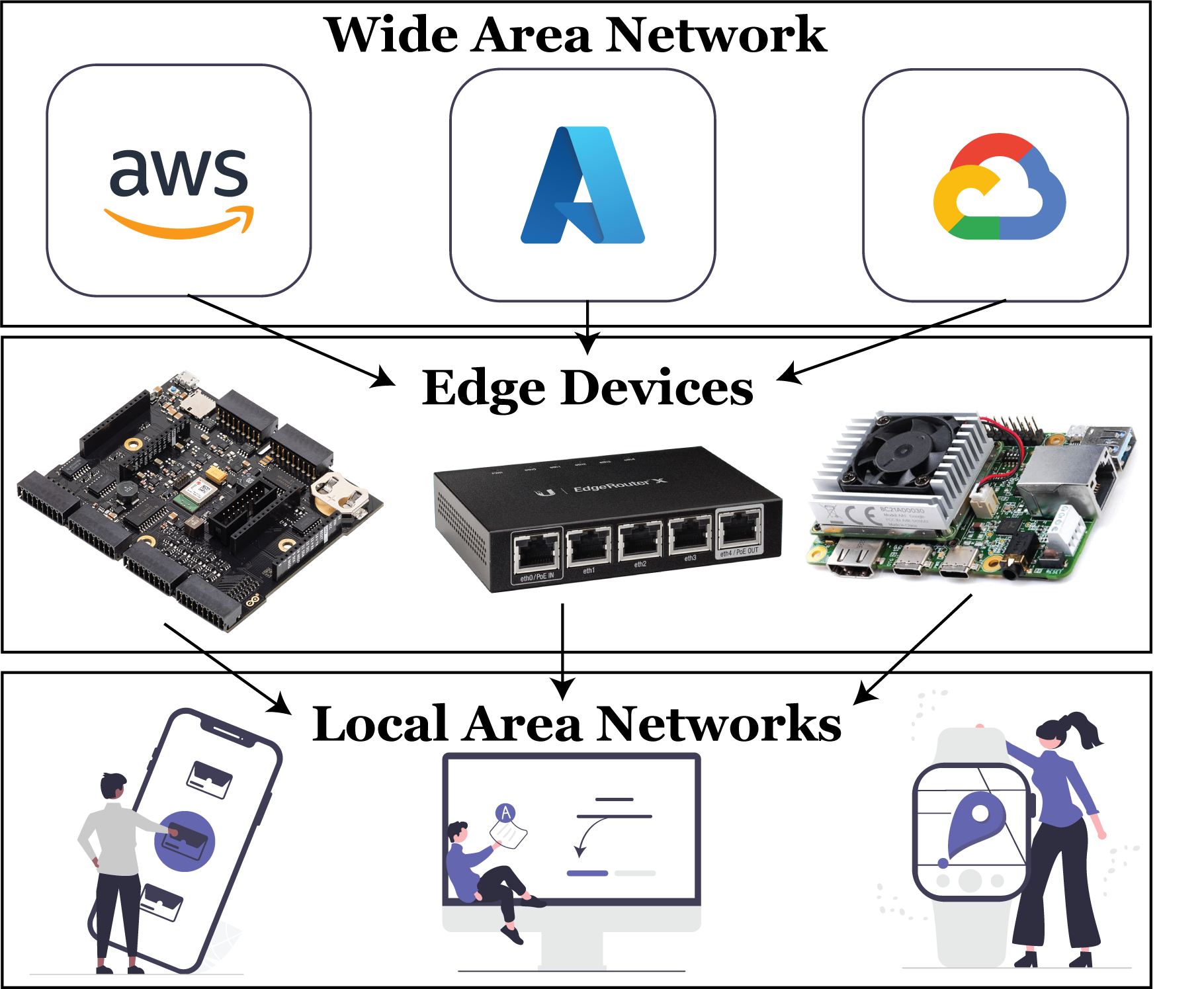}
    \caption{\footnotesize {\bf What is an edge device } }
    \vspace{-3mm}
    \label{fig:edge}
\end{figure}
\subsection*{Intro to EdgeImpulse}

\textbf{EdgeImpulse}\cite{hymel_edge_2022} is a development platform for machine learning on edge. 
EdgeImpulse is working to take the complicated portions of a computer science (CS) idea and make it digestible to non-CS audiences through a simple online web interface. 

\subsection*{Using Edge Impulse to Connect with SDGs}
In 2015, the United Nations proposed 17 Sustainable Development Goals (SDGs) to address environmental, political, and economic challenges on a global scale\cite{gostin_sustainable_2015}. Each goal highlights a broad challenge to sustainability broken down into specific targets designed to draw attention to and proactively pursue solutions. In recent years, the United Nations has worked to develop pedagogical resources associated with the SDGs to make them more accessible to academia.  These resources include teaching guides, learning outcomes, and evaluative assessments\cite{tang_education_2017}.  In addition to teaching resources, the SDGs rely on innovative solutions from engineers and effective education practices to train the next generation of sustainability-minded communities\cite{el-jardali_changing_2018}. 

The cross between new and advancing technologies of both hardware and software with fields like ecology has become known as conservation technology (CT) or Tech4Wildlife\cite{bergertal_conservation_2018}. Much like the computational science of AI4Good, the goal of CT is to apply these techniques to humanitarian solutions and conservation\cite{schulz_conservation_2023}. The field of conservation technology does not currently have a curriculum framework in any of the fields of engineering, computer science, or biology. Still, we have been working to develop some foundational frameworks, including learning objectives\cite{schulz_foundational_2022} and thematic elements of the course that make it successful\cite{schulz_towards_2022}. 

Edge Impulse, in its framework of being a learning and teaching tool in the field of computer vision and machine learning, has several case studies that have been implemented in classrooms. We highlight a few of these case studies, specifically focused on introducing the EdgeImpulse tool-kits in an Environmental Engineering classroom. 

Educators can sign up for ten free kits for Edge Impulse's educator program to provide edge devices in engineering classrooms. This could in turn help connect specific ideas of advanced computing techniques to engineering education and other fields such as environmental or bio-systems engineering.

\section{Engineering Education Projects for the Classroom}
We will discuss four case studies to bring into the classroom for various engineering disciplines and items in the environmental and sustainability engineering space. Each case study has online how-to guides, linked throughout this paper, and could be utilized in environmental or sustainability engineering curricula, classrooms, or laboratory teaching settings. Each case study targets specific ABET criteria for environmental engineering programs\cite{abet_criteria_2019}. Many of these projects would require purchasing some hardware components for a lab setting, and we will discuss each of those in the respective sections as well as successful applications and real-world examples that have utilized these types of techniques. 

\subsection{Air Quality Detection Using Accelerometer}
Currently, air quality is a good marker of engineering for sustainable development given the direct connections with climate change, climate action, and data analysis\cite{mazutti_smart_2020}. An advantage of air quality is that data acquisition and estimations of air quality can be obtained with simple sensors and readings. A development board and a three-axis accelerometer can be combined with Edge Impulse studio to detect different types of air quality\cite{bild_must_2022}. As with each of the case studies we discuss here, the advantage of this device is the training data can be collected relatively simply with an air purifier in a laboratory setting. A labeled training data set is required for all computer vision projects discussed in this paper. The classification for this air quality detection project is tested through four different conditions: clean air, slightly polluted, highly polluted, and controlled. There are air purifiers in many classrooms now due to COVID-19 protection at universities, and they are a widely accessible way to generate a training set. 

\noindent\textbf{Limitations $\&$ Opportunities:} Overall, this is the most accessible case study we will discuss as there are no student limitations regarding access/security/etc. Additionally, this project provides opportunities to access and include additional data, such as looking at the air in different buildings across campus. This could allow students to create a data set for tracking differences in air quality between different air filtration systems through the centuries, since campuses often have buildings predating modern filtration systems. 

\subsection{Roadside Litter Detector}
Another application in an environmental engineering classroom is non-sustainability with current waste treatment plans. Littering is a significant issue that introduces the human element directly into the human-caused environmental challenges and has connections with the sustainable development goals regarding SDG 6 of sanitation. Over 11 billion dollars in the United States is spent yearly to clean up litter. A significant challenge with litter on places like highways is the inability to assess where the litter is located. Nathaniel Felleke has worked on solving this using an edge device and the edge impulse online tools through a roadside liter detector\cite{bild_cleanin_2022}. 

By combining a raspberry pi four board, a wireless notecard, and a computer webcam while driving along the highway, he generated a training data set to identify locations where liter is present on the side of the highway as a coin flip distinction of trash versus no-trash. Using the Edge Impulse Studio, further analysis could be completed, and the image accuracy for this data set had almost a 90$\%$ accuracy rate. 

\noindent\textbf{Limitations $\&$ Opportunities:} Overall, the hardware components of this exercise are cheap and often already in engineering classrooms with items like raspberry pi boards interfacing with many different types of laboratory experiments in modern engineering education, but this specific project requires cars. Alternatively, the same methods could be performed with a system mounted onto a bike or even on the backpack of a student walking on campus to look at trash and litter on the side of the road or sidewalk. This requires some basic knowledge of coding and interfacing between some devices and would be a more rigorous exercise, but it has a lot of engineering connections.

\subsection{Automated Bird Identifier}
The third case study focuses on combining edge devices with allowing students to understand the field of community science, which is an ever-increasing field for large data collection\cite{johnston_analytical_2021}. For years, birders and wildlife experts have used bird books and field guides to identify birds from afar. Merlin Bird ID is an application designed to eliminate the need for expertise and expand birding to all levels of bird watchers. Understanding that birding may be inherently remote, the Merlin application downloads the machine learning models and data required onto any iOS or Android device. It then can run entirely on users' mobile devices without an internet connection. As described above, this makes Merlin Bird ID an example of an edge application. 

Merlin identifies birds by asking users for photos of birds and audio from birds' calls, as well as the time and location. Its machine-learning model combines the birding community's 800 million sightings on eBird with annotations from experts at Cornell's Lab of Ornithology. This model is distilled into something small enough to fit onto a mobile device with minimal space to reach the mass public. Items like eBird have become a successful example of community-driven scientific data collection\cite{sullivan_ebird_2014}. For those who want to work on interfacing their hardware setup, there are ways to do this using either the Lacuna space and things network or a more open-source toolkit such as the AudioMoth that allows for the classification of bird species using sound \cite{lequertier_bird_2021}. There are online tutorials that take you through constructing your device to determine accuracy for distinguishing between different bird species, such as a house sparrow and parakeet, where they give the example of an accuracy of 91 $\%$, with a 9 $\%$ of incorrect species differentiation. Following the build guide, students would be able to interface and learn about more than just the ability of edge devices to use images but also data in the form of sounds. 

\noindent\textbf{Limitations $\&$ Opportunities:} This project also shows students the broader connection to current data and programs, with items like Merlin application, iNaturalist, and eBird having data that is purely driven by community scientists' involvement in the collection of data sets. Additionally, when training data sets for this, there is a wide range of open-source data sets and sounds of pre-identified birds in various regions of the globe. Students can also put one of these devices outside their dorm window to see which bird species are around campus. 

\subsection{Wildlife Camera Trap Detection}
\begin{figure}
\centering
\includegraphics[width=1\textwidth]{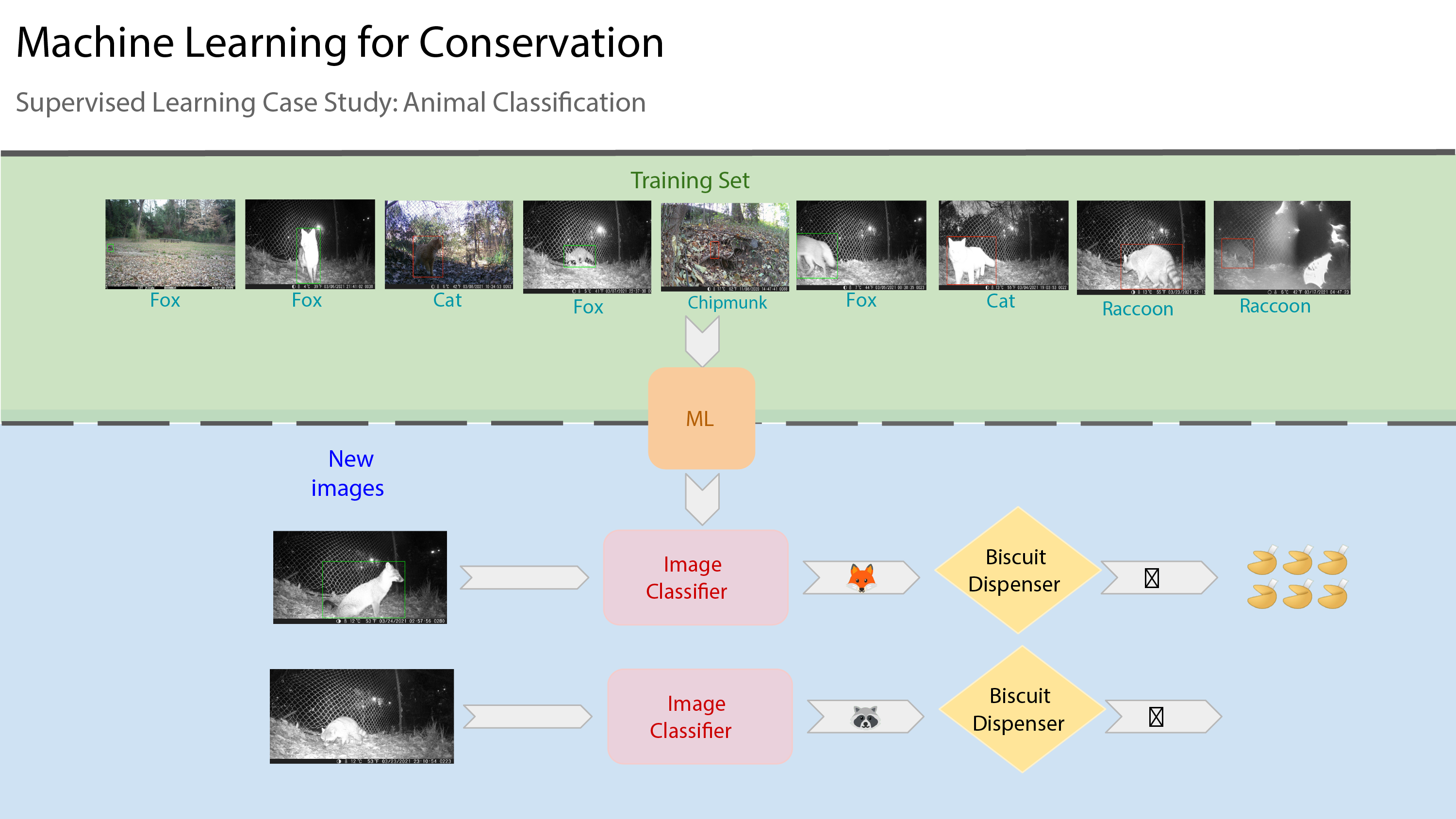}
    \caption{\footnotesize {\bf Example of machine learning training set imposed on differentiation between fox and not fox } }
    \vspace{-3mm}
    \label{fig:ml_conservation}
\end{figure}

Finally, the fourth iteration would be an example project that is related to the use of camera trap data to distinguish between specific species. This idea is well known in the field of conservation technology, discussed in the introduction in different ways. This includes using computer vision to identify between, say, a fox versus not a fox, or even can get down to the identification of specific fox individuals (\fig{fig:ml_conservation}). 
Different wildlife cameras are currently being developed that specify target species, and there are many use cases in literature utilizing edge devices. We will give one example here with that of EleTect\cite{sheth_eletect_nodate}. This technique and application can be applied to other camera-collected data, and identification between, say, fox versus not a fox or elephant versus not an elephant can be a simple task. There are specific animals around campuses or student dorms that would provide plentiful opportunities to explore this case study as a laboratory project. 

\noindent\textbf{Limitations $\&$ Opportunities:} The primary challenge or limitation of this case study is security when looking at including camera trap information that could be taken around campus. Additionally, many cameras that are placed around campus will need to have tags on them for security reasons. Therefore before implementing any data collection on or around campus, it is recommended to check with your research ethics departments and campus police and groundskeepers to get the best amount of data. Additionally, this allows students to begin with a simple fox versus not-fox idea and expand to individual species identification using the Edge Impulse online tools. 

\section{Discussion}
\subsection{Connections between Machine Learning and ABET Outcomes}
These case studies and Machine learning align directly with several ABET outcomes. Primarily when we view the Environmental Engineering ABET outcomes, we will discuss each outcome and the relevance to Machine Learning and how they are connected through the case studies discussed in the previous section. 
\begin{itemize}
    \item \textbf{ABET Outcome:} Mathematics through differential equations, probability and statistics, calculus-based physics, chemistry (including stoichiometry, equilibrium, and kinetics), earth science, biological science, and fluid mechanics.\\ \textbf{Relevance to Machine Learning:} ML as a field utilizes advanced techniques of statistics and mathematics that are novel and new computing techniques, as discussed in the introduction. 
\item  \textbf{ABET Outcome:} Material and energy balances; fate and transport of substances in and between air, water, and soil phases; and advanced principles and practices relevant to the program objectives. - \textbf{Relevance to Machine Learning:} Discussed in the case studies above are direct links towards using these case studies to view and understand items about the air, soil, and other environmental factors. 
\item  \textbf{ABET Outcome:} Hands-on laboratory experiments and analysis and interpretation of the resulting data in more than one central environmental engineering focus area, e.g., air, water, land, and environmental health. \\ \textbf{Relevance to Machine Learning:} Each case study gives an example case for students to utilize a device like the Edge Impulse board to generate a laboratory experiment that allows actual testing of an analysis method in real-time. 
\item  \textbf{ABET Outcome:} Design of environmental engineering systems considering risk, uncertainty, sustainability, life-cycle principles, and environmental impacts. - \textbf{Relevance to Machine Learning:} Machine Learning has risks, biases, and other challenges to overcome with these different types of problems and solutions, which are suitable to teach students before they are in the workforce. 
\end{itemize}

\subsection*{Challenges in Educating ML to Non-Computer Scientists}

Teaching machine learning to an interdisciplinary, non-computer science crowd has its challenges. There is existing literature about teaching machine learning, deep learning, or computer vision to those with a background in computer science or a related discipline.\cite{shouman2022experiences, steinbach2021teaching, spurlock2017making}. And other works concentrate on teaching machine learning to non-computer scientists, including undergraduates \cite{sulmont_what_2019}, business majors \cite{wunderlich_machine_2021}, artists \cite{fiebrink_machine_2019}, material science engineers \cite{sun_teaching_2022}, biologists \cite{magnano_approachable_2022}, and ecologists \cite{cole_cv4ecology}. 
\clearpage
However, common challenges faced when teaching machine learning to groups of non-computer science students included the following:

\begin{enumerate}
    \item Designing sufficient coding structure and guidelines. Computer science students tend to have personal preferences in their setups; some might prefer coding on sublime and running their code via the command line, while others might prefer coding in what's known as an integrated development environment (IDE). This difference is akin to, though it has more variance than, using Microsoft Word versus text editing on a PC. Because of computer scientists' different personal preferences, it is not standard for these types of setup details to be included or required in machine learning classes. Non-computer scientists, however, may need more structure to ensure that they can collaborate on solving common coding problems and streamline the learning process.
    \item Removing ambiguity from infrastructure and process recommendations for machine learning work. If a class is structured to run on a virtual machine (VM), for example, there are several ways in which someone could get their code on their VM. They could write code locally and copy it directly to their VM using \texttt{rsync} or \texttt{scp}. Alternatively, they could write code locally, \texttt{push} their code to a GitHub (or comparable) repository, and then have their VM \texttt{pull} code from the remote repository. Computer science students may have a preferred workflow, but non-computer science students may benefit from more than a single structured recommendation.
    \item Avoiding wrappers to existing machine learning libraries. ``Wrappers" are designed to make their underlying libraries more accessible or digestible. However, some have found that using them in an educational setting makes them difficult to inspect or debug when faced with a challenge. They hide a lot of complexity, which seems wonderful in practice, but makes learning about what is going on under the hood much more difficult.\cite{cole_cv4ecology}
    \item Access to more powerful devices. Computer science students may have access to more powerful machines, such as a school's GPU-computing cluster. Non-computer science students or people outside academia trying to learn machine learning independently may not have such access. Cloud services like Microsoft's Azure and Amazon's AWS frequently provide free computing credits for educational purposes, though those credits may only extend to lower-power (less costly) devices. As a result, required access to more powerful devices may financially limit the public from being able to run on powerful infrastructure.\cite{cole_cv4ecology}
\end{enumerate}

Distilling structure and standardizing processes streamline workflows and give a joint base for non-computer-science students and the public to launch. With these defined, machine learning education becomes more standardized and digestible to anyone attempting to learn.

\subsubsection*{Identifying bias in computer vision models}
In machine learning, bias indicates the difference between an ML model's average prediction and the correct value. High bias in a machine learning model indicates oversimplification, whereas low bias indicates that a machine learning model might be too specifically fit on the training data.

For this paper, however, we define bias as the \textit{human bias} that goes into producing a machine learning system. Bias can be introduced to a system in several forms. It can be encoded in the training system by the data fed into training (selection bias): for example, if only mugshots are used to train a face recognition system, the machine learning model "biases" toward populations that are more often incarcerated. Alternatively, bias can be introduced in the algorithm itself. Continuing with the mugshot example, this could happen if the person labeling the photos adds their stereotypes to the labels. 

Ongoing research is trying to point out and mitigate these innate biases in facial recognition. Recent works, for example, have successfully predicted the performance of face-identification models without labeled data, removing the bias introduced from labeling and allowing face-recognition software to test their models' biases. \cite{dengUnsupervicedFRECCV2022, sernaICPR2021}

\section{Future Work}
In the future, we are going to perform a more formalized understanding of the types of content and identities that are gained when engineering students use edge devices in the classroom. Machine learning and computer vision are becoming more applicable to all types of engineering careers in both industry and academia. It should be a priority in the coming years to educate undergraduate students to use these techniques and tools. In the future, we will also look at utilizing edge devices outside of just the field of environmental engineering and determine the impact these lessons can have on the engineering curriculum to help educate the next generation of computer vision-minded engineers using edge devices. 

\footnotesize 

\bibliographystyle{unsrt} 

\begin{thebibliography}{10}

\bibitem{unsdgs}
\url{https://sdgs.un.org/goals}.

\bibitem{sdgs2020}
Azizpour H. Leite I. et~al. Vinuesa, R.
\newblock The role of artificial intelligence in achieving the sustainable
  development goals.
\newblock {\em Nature Communications}, 11(233), 2020.

\bibitem{acemoglu2018}
D.~Acemoglu and P.~Restrepo.
\newblock Artificial intelligence, automation, and work.
\newblock {\em National Bureau of Economic Research}, Working Paper(24196),
  2018.

\bibitem{bolukbasi2016}
Chang K.-W. Zou J. Saligrama~V. Bolukbasi, T. and A.~Kalai.
\newblock Man is to computer programmer as woman is to homemaker?
\newblock {\em Adv. Neural Inf. Process. Syst.}, 29:4349--4357, 2016.

\bibitem{norouzzadeh2018}
M.~S. et~al. Norouzzadeh.
\newblock Automatically identifying, counting, and describing wild animals in
  camera-trap images with deep learning.
\newblock {\em Proc. Natl Acad. Sci. USA}, 115:E5716--E5725, 2018.

\bibitem{un2015}
UN~General~Assembly (UNGA).
\newblock A/res/70/1 transforming our world: the 2030 agenda for sustainable
  development.
\newblock {\em Resolut}, 25:1--35, 2015.

\bibitem{cs221GeneralAISlides}
Percy Liang and Dorsa Sadigh.
\newblock Ai general history, 2022.


\bibitem{MostafaLFD2012}
Yaser~S. Abu-Mostafa, Malik Magdon-Ismail, and Hsuan-Tien Lin.
\newblock {\em Learning From Data}.
\newblock AMLBook, 2012.

\bibitem{de_freitas_im_2021}
Adrian~A. de~Freitas and Troy~B. Weingart.
\newblock I'm {Going} to {Learn} {What}?!? {Teaching} {Artificial}
  {Intelligence} to {Freshmen} in an {Introductory} {Computer} {Science}
  {Course}.
\newblock In {\em Proceedings of the 52nd {ACM} {Technical} {Symposium} on
  {Computer} {Science} {Education}}, {SIGCSE} '21, pages 198--204, New York,
  NY, USA, March 2021. Association for Computing Machinery.

\bibitem{steinbach_teaching_2021}
Peter Steinbach, Heidi Seibold, and Oliver Guhr.
\newblock Teaching {Machine} {Learning} in 2020.
\newblock In {\em Proceedings of the {First} {Teaching} {Machine} {Learning}
  and {Artificial} {Intelligence} {Workshop}}, pages 1--6. PMLR, March 2021.
\newblock ISSN: 2640-3498.

\bibitem{shouman_experiences_2022}
Omar Shouman, Simon Fuchs, and Holger Wittges.
\newblock Experiences from {Teaching} {Practical} {Machine} {Learning}
  {Courses} to {Master}’s {Students} with {Mixed} {Backgrounds}.
\newblock In {\em Proceedings of the {Second} {Teaching} {Machine} {Learning}
  and {Artificial} {Intelligence} {Workshop}}, pages 62--67. PMLR, March 2022.
\newblock ISSN: 2640-3498.

\bibitem{cole_teaching_2023}
Elijah Cole, Suzanne Stathatos, Björn Lütjens, Tarun Sharma, Justin Kay,
  Jason Parham, Benjamin Kellenberger, and Sara Beery.
\newblock Teaching {Computer} {Vision} for {Ecology}, January 2023.
\newblock arXiv:2301.02211 [cs].

\bibitem{lauer_multi-animal_2022}
Jessy Lauer, Mu~Zhou, Shaokai Ye, William Menegas, Steffen Schneider, Tanmay
  Nath, Mohammed~Mostafizur Rahman, Valentina Di~Santo, Daniel Soberanes,
  Guoping Feng, Venkatesh~N. Murthy, George Lauder, Catherine Dulac,
  Mackenzie~Weygandt Mathis, and Alexander Mathis.
\newblock Multi-animal pose estimation, identification and tracking with
  {DeepLabCut}.
\newblock {\em Nature Methods}, 19(4):496--504, April 2022.
\newblock Number: 4 Publisher: Nature Publishing Group.

\bibitem{mooney_big_2018}
Stephen~J. Mooney and Vikas Pejaver.
\newblock Big {Data} in {Public} {Health}: {Terminology}, {Machine} {Learning},
  and {Privacy}.
\newblock {\em Annual Review of Public Health}, 39(1):95--112, 2018.
\newblock \_eprint: https://doi.org/10.1146/annurev-publhealth-040617-014208.

\bibitem{tuia_perspectives_2022}
Devis Tuia, Benjamin Kellenberger, Sara Beery, Blair~R. Costelloe, Silvia
  Zuffi, Benjamin Risse, Alexander Mathis, Mackenzie~W. Mathis, Frank van
  Langevelde, Tilo Burghardt, Roland Kays, Holger Klinck, Martin Wikelski,
  Iain~D. Couzin, Grant van Horn, Margaret~C. Crofoot, Charles~V. Stewart, and
  Tanya Berger-Wolf.
\newblock Perspectives in machine learning for wildlife conservation.
\newblock {\em Nature Communications}, 13(1):792, February 2022.
\newblock Number: 1 Publisher: Nature Publishing Group.

\bibitem{ackermann_deploying_2018}
Klaus Ackermann, Joe Walsh, Adolfo De~Unánue, Hareem Naveed, Andrea
  Navarrete~Rivera, Sun-Joo Lee, Jason Bennett, Michael Defoe, Crystal Cody,
  Lauren Haynes, and Rayid Ghani.
\newblock Deploying {Machine} {Learning} {Models} for {Public} {Policy}: {A}
  {Framework}.
\newblock In {\em Proceedings of the 24th {ACM} {SIGKDD} {International}
  {Conference} on {Knowledge} {Discovery} \& {Data} {Mining}}, {KDD} '18, pages
  15--22, New York, NY, USA, July 2018. Association for Computing Machinery.

\bibitem{khanAcademy}
Khan academy.
\newblock \url{https://www.khanacademy.org/}.

\bibitem{coursera}
Coursera.
\newblock \url{https://www.coursera.com/}.

\bibitem{udacity}
Udacity.
\newblock \url{https://www.udacity.com/}.

\bibitem{aws}
Amazon aws in education.
\newblock \url{https://aws.amazon.com/education/ml-in-education/}.

\bibitem{gcp}
Google gcp in education.
\newblock \url{https://cloud.google.com/edu}.

\bibitem{azure}
Microsoft azure for education.
\newblock \url{https://azureforeducation.microsoft.com/en-us/Institutions}.

\bibitem{linkedin_news_linkedin_2022}
LinkedIn News.
\newblock {LinkedIn} {Jobs} on the {Rise} 2022: {The} 25 {U}.{S}. roles that
  are growing in demand.
\newblock Technical report, 2022.

\bibitem{overthewall}
Jonathan Owens and Rachel Cooper.
\newblock The importance of a structured new product development (npd) process:
  A methodology.
\newblock pages 10/1--10/6, 02 2001.

\bibitem{weerts_teaching_2022}
Hilde Jacoba~Petronella Weerts and Mykola Pechenizkiy.
\newblock Teaching {Responsible} {Machine} {Learning} to {Engineers}.
\newblock In {\em Proceedings of the {Second} {Teaching} {Machine} {Learning}
  and {Artificial} {Intelligence} {Workshop}}, pages 40--45. PMLR, March 2022.
\newblock ISSN: 2640-3498.

\bibitem{sulmont_what_2019}
Elisabeth Sulmont, Elizabeth Patitsas, and Jeremy~R Cooperstock.
\newblock What is hard about teaching machine learning to non-majors?
  {Insights} from classifying instructors’ learning goals.
\newblock {\em ACM Transactions on Computing Education (TOCE)}, 19(4):1--16,
  2019.
\newblock Publisher: ACM New York, NY, USA.

\bibitem{wunderlich_machine_2021}
Linus Wunderlich, Allen Higgins, and Yossi Lichtenstein.
\newblock Machine {Learning} for {Business} {Students}: {An} {Experiential}
  {Learning} {Approach}.
\newblock In {\em Proceedings of the 26th {ACM} {Conference} on {Innovation}
  and {Technology} in {Computer} {Science} {Education} {V}. 1}, pages 512--518,
  2021.

\bibitem{fiebrink_machine_2019}
Rebecca Fiebrink.
\newblock Machine learning education for artists, musicians, and other creative
  practitioners.
\newblock {\em ACM Transactions on Computing Education (TOCE)}, 19(4):1--32,
  2019.
\newblock Publisher: ACM New York, NY, USA.

\bibitem{sun_teaching_2022}
Shijing Sun, Keith Brown, and A~Gilad Kusne.
\newblock Teaching machine learning to materials scientists: {Lessons} from
  hosting tutorials and competitions.
\newblock {\em Matter}, 5(6):1620--1622, 2022.
\newblock Publisher: Elsevier.

\bibitem{magnano_approachable_2022}
Chris~S Magnano, Fangzhou Mu, Rosemary~S Russ, Milica Cvetkovic, Debora Treu,
  and Anthony Gitter.
\newblock An approachable, flexible, and practical machine learning workshop
  for biologists.
\newblock {\em bioRxiv}, 2022.
\newblock Publisher: Cold Spring Harbor Laboratory.

\bibitem{hymel_edge_2022}
Shawn Hymel, Colby Banbury, Daniel Situnayake, Alex Elium, Carl Ward, Mat
  Kelcey, Mathijs Baaijens, Mateusz Majchrzycki, Jenny Plunkett, David
  Tischler, Alessandro Grande, Louis Moreau, Dmitry Maslov, Artie Beavis, Jan
  Jongboom, and Vijay~Janapa Reddi.
\newblock Edge {Impulse}: {An} {MLOps} {Platform} for {Tiny} {Machine}
  {Learning}, November 2022.
\newblock arXiv:2212.03332 [cs].

\bibitem{gostin_sustainable_2015}
Lawrence~O. Gostin and Eric~A. Friedman.
\newblock The {Sustainable} {Development} {Goals}: {One}-{Health} in the
  {World}'s {Development} {Agenda}.
\newblock {\em JAMA}, 314(24):2621--2622, December 2015.

\bibitem{tang_education_2017}
Qian Tang.
\newblock {\em Education for {Sustainable} {Development} {Goals}: {Learning}
  {Objectives}}.
\newblock United Nations Educational, Scientific, and Cultural Organization,
  2017.

\bibitem{el-jardali_changing_2018}
Fadi El-Jardali, Nour Ataya, and Racha Fadlallah.
\newblock Changing roles of universities in the era of {SDGs}: rising up to the
  global challenge through institutionalising partnerships with governments and
  communities.
\newblock {\em Health Research Policy and Systems}, 16(1):38, May 2018.

\bibitem{bergertal_conservation_2018}
Oded Berger tal, and Jose J. Lahoz Monfort.
\newblock Conservation technology: {The} next generation.
\newblock {\em Conservation Letters}, 11(6):e12458, 2018.

\bibitem{schulz_conservation_2023}
Andrew Schulz, Cassie Shriver, Suzanne Stathatos, Benjamin Seleb, Emily Weigel,
  Young-Hui Chang, M.~Saad Bhamla, David Hu, and Joseph~R. Mendelson~III.
\newblock Conservation {Tools}: {The} {Next} {Generation} of
  {Engineering}--{Biology} {Collaborations}, January 2023.
\newblock arXiv:2301.01103 [cs, q-bio].

\bibitem{schulz_foundational_2022}
Andrew Schulz, Anika Patka, Cassandra Shriver, Margaret Zhang, Nima Jadali,
  D.~L. Hu, and Roxanne Moore.
\newblock A {Foundational} {Design} {Experience} in {Conservation}
  {Technology}: {A} {Multi}-{Disciplinary} {Approach} to {Meeting}
  {Sustainable} {Development} {Goals}.
\newblock In {\em American {Society} of {Engineering} {Education}}. ASEE
  Conferences, June 2022.

\bibitem{schulz_towards_2022}
Andrew Schulz, Caroline Greiner, Benjamin Seleb, Cassandra Shriver, D.~L. Hu,
  and Roxanne Moore.
\newblock Towards the {UN}'s {Sustainable} {Development} {Goals} ({SDGs}):
  {Conservation} {Technology} for {Design} {Teaching} \& {Learning}.
\newblock In {\em American {Society} of {Engineering} {Education}}, March 2022.

\bibitem{abet_criteria_2019}
ABET.
\newblock Criteria for {Accrediting} {Engineering} {Programs}, 2019 – 2020
  {\textbar} {ABET}.
\newblock Technical report, 2019.

\bibitem{mazutti_smart_2020}
Janaina Mazutti, Luciana Londero~Brandli, Amanda Lange~Salvia, Bárbara~Maria
  Fritzen~Gomes, Luana~Inês Damke, Vanessa Tibola~da Rocha, and Roberto~dos
  Santos~Rabello.
\newblock Smart and learning campus as living lab to foster education for
  sustainable development: an experience with air quality monitoring.
\newblock {\em International Journal of Sustainability in Higher Education},
  21(7):1311--1330, January 2020.
\newblock Publisher: Emerald Publishing Limited.

\bibitem{bild_must_2022}
Nick Bild.
\newblock Must {Be} {Something} in the {Air}, September 2022.

\bibitem{bild_cleanin_2022}
Nick Bild.
\newblock Cleanin' {Up} the {Town} with a {Litter} {Heatmap}, June 2022.

\bibitem{johnston_analytical_2021}
Alison Johnston, Wesley~M. Hochachka, Matthew~E. Strimas-Mackey, Viviana
  Ruiz~Gutierrez, Orin~J. Robinson, Eliot~T. Miller, Tom Auer, Steve~T.
  Kelling, and Daniel Fink.
\newblock Analytical guidelines to increase the value of community science
  data: {An} example using {eBird} data to estimate species distributions.
\newblock {\em Diversity and Distributions}, 27(7):1265--1277, 2021.
\newblock \_eprint: https://onlinelibrary.wiley.com/doi/pdf/10.1111/ddi.13271.

\bibitem{sullivan_ebird_2014}
Brian~L. Sullivan, Jocelyn~L. Aycrigg, Jessie~H. Barry, Rick~E. Bonney,
  Nicholas Bruns, Caren~B. Cooper, Theo Damoulas, André~A. Dhondt, Tom
  Dietterich, Andrew Farnsworth, Daniel Fink, John~W. Fitzpatrick, Thomas
  Fredericks, Jeff Gerbracht, Carla Gomes, Wesley~M. Hochachka, Marshall~J.
  Iliff, Carl Lagoze, Frank~A. La~Sorte, Matthew Merrifield, Will Morris,
  Tina~B. Phillips, Mark Reynolds, Amanda~D. Rodewald, Kenneth~V. Rosenberg,
  Nancy~M. Trautmann, Andrea Wiggins, David~W. Winkler, Weng-Keen Wong,
  Christopher~L. Wood, Jun Yu, and Steve Kelling.
\newblock The {eBird} enterprise: {An} integrated approach to development and
  application of citizen science.
\newblock {\em Biological Conservation}, 169:31--40, January 2014.

\bibitem{lequertier_bird_2021}
Aurelien Lequertier, Jenny Plunkett, and Raul James.
\newblock Bird {Classification} in {Remote} {Areas} with {Lacuna} {Space} and
  {The} {Things} {Network}, January 2021.

\bibitem{sheth_eletect_nodate}
Dhruv Sheth.
\newblock {EleTect} - {TinyML} and {IoT} {Based} {Smart} {Wildlife} {Tracker}.

\bibitem{shouman2022experiences}
Omar Shouman, Simon Fuchs, and Holger Wittges.
\newblock Experiences from teaching practical machine learning courses to
  master’s students with mixed backgrounds.
\newblock In {\em Proceedings of the Second Teaching Machine Learning and
  Artificial Intelligence Workshop}, pages 62--67. PMLR, 2022.

\bibitem{steinbach2021teaching}
Peter Steinbach, Heidi Seibold, and Oliver Guhr.
\newblock Teaching machine learning in 2020.
\newblock In {\em European Conference on Machine Learning and Principles and
  Practice of Knowledge Discovery in Databases}, pages 1--6. PMLR, 2021.

\bibitem{spurlock2017making}
Scott Spurlock and Shannon Duvall.
\newblock Making computer vision accessible for undergraduates.
\newblock {\em Journal of Computing Sciences in Colleges}, 33(2):215--221,
  2017.

\bibitem{cole_cv4ecology}
Elijah Cole, Suzanne Stathatos, Björn Lütjens, Tarun Sharma, Justin Kay,
  Jason Parham, Benjamin Kellenberger, and Sara Beery.
\newblock Teaching computer vision for ecology, 2023.

\bibitem{dengUnsupervicedFRECCV2022}
Alexandra Chouldechova, Siqi Deng, Yongxin Wang, Wei Xia, and Pietro Perona.
\newblock Unsupervised and semi-supervised bias benchmarking in face
  recognition.
\newblock In Shai Avidan, Gabriel Brostow, Moustapha Ciss{\'e}, Giovanni~Maria
  Farinella, and Tal Hassner, editors, {\em Computer Vision -- ECCV 2022},
  pages 289--306. Springer Nature Switzerland, 2022.

\bibitem{sernaICPR2021}
Ignacio Serna, Alejandro Peña, Aythami Morales, and Julian Fierrez.
\newblock Insidebias: Measuring bias in deep networks and application to face
  gender biometrics.
\newblock In {\em 2020 25th International Conference on Pattern Recognition
  (ICPR)}, pages 3720--3727, 2021.

\end{thebibliography}


\end{document}